Spin and orbital magnetic moments in perpendicularly magnetized $Ni_{1-x}Co_{2+y}O_{4-z}$ epitaxial thin films: Effects of site-dependent cation valence states

Daisuke Kan[1,a], Masaichiro Mizumaki[2], Miho Kitamura[3], Yoshinori Kotani[2], Yufan Shen[1], Ikumi Suzuki[1], Koji Horiba[3], Yuichi Shimakawa[1]

[1]Institute for Chemical Research, Kyoto University, Uji, Kyoto 611-0011, Japan

[2]Japan Synchrotron Radiation Research Institute, SPring-8, Sayo, Hyogo 679-5198, Japan

[3]Photon Factory, Institute of Materials Structure Science, High Energy Accelerator Research Organization (KEK), Tsukuba 305-0801, Japan.

[a] Electronic mail : dkan@scl.kyoto-u.ac.jp

We carried out x-ray absorption spectroscopy (XAS) and x-ray magnetic circular dichroism (XMCD) spectroscopy and investigated cation valence states and spin and orbital magnetic moments in the inverse-spinel ferrimagnet $Ni_{1-x}Co_{2+y}O_{4-z}$ (NCO) epitaxial films with the perpendicular magnetic anisotropy. We show that the oxygen pressure $P_{O2}$ during the film growth by pulsed laser deposition influences not only the cation stoichiometry (site-occupation) but also the cation valence state. Our XAS results show that the Ni in the $O_h$-site is in the intermediate valence state between +2 and +3, $Ni^{(2+\delta)+}$ ($0<\delta<1$), whose nominal valence state (the $\delta$ value) varies depending on $P_{O2}$. On the other hand, the Co in the octahedral ($O_h$) and tetrahedral ($T_d$) sites respectively have the valence state close to +3 and +2. We also find that the XMCD signals originate mainly from the $T_d$-site $Co^{2+}$ (Co_Td) and $O_h$-site $Ni^{(2+\delta)+}$ (Ni_Oh), indicating that these cation valence states are the key in determining the magnetic and transport properties of NCO films. Interestingly, the valence state of $Ni^{(2+\delta)+}$ that gives rise to the XMCD signal remains unchanged independent of $P_{O2}$. The electronic structure of $Ni^{(2+\delta)+}$ that is responsible for the magnetic moment and electrical conduction differs from those of $Ni^{2+}$ and $Ni^{3+}$. In addition, the orbital magnetic moment originating from Co_Td is as large as 0.14 $\mu_B$/Co_Td and parallel to the magnetization while the Ni_Oh orbital moment is as small as 0.07 $\mu_B$/Ni_Oh and is rather isotropic. The Co_Td therefore plays the key role in the perpendicular magnetic anisotropy of the films. Our results demonstrate the significance of the site-dependent cations valence states for the magnetic and transport properties of $NiCo_2O_4$ films.

I. Introduction

Transition metals in oxides can have various valence states through orbital hybridization with oxygen, impacting various functional properties. For ternary spinel oxides, transition metals occupying tetrahedral and octahedral sites can have different valence states, leading to a variety of transport and magnetic properties. Evaluating cation valence states in spinel oxides and delineating their influence on physical properties are therefore important.

The inverse spinel oxide $NiCo_2O_4$ [1-5] in which Co is in both tetrahedral ($T_d$) and octahedral ($O_h$) sites while Ni is in only the $O_h$ site, has been shown to have a variety of its properties, such as above-room-temperature ferrimagnetism[3,5], metallic electrical conduction[6-8], half-metallic properties[6,7], perpendicular magnetic anisotropy[9,10], and electrochemical activities [11-14]. These findings have revealed the potential application of this oxide for spintronic and electrochemical devices. Therefore, delineating correlations between cation valence states and functional properties in $NiCo_2O_4$ is crucial. Recently it has been shown that transport and magnetic properties of epitaxial films of this oxide strongly depend on their growth conditions[6,15-18]. Our resonant x-ray diffraction measurements[19] for films grown by pulsed laser deposition have also revealed that cation distribution ($T_d$- and $O_h$-site occupation of Co and Ni) in $NiCo_2O_4$ films vary depending on oxygen partial pressures ($P_{O_2}$s) during film growth. Given that Co and Ni could accommodate various valence states (+2 and +3, for example), growth conditions such as $P_{O_2}$ would influence not only compositions but also valence states of Co and Ni, affecting films' properties. It is therefore interesting to see correlations between cation valence states and physical properties of $NiCo_2O_4$ films grown under various $P_{O_2}$s.

In this study, we carried out x-ray absorption spectroscopy (XAS) and x-ray magnetic circular dichroism (XMCD) spectroscopy and investigated valence states and spin and orbital magnetic moments of Co and Ni in $Ni_{1-x}Co_{2+y}O_{4-z}$ (NCO) epitaxial films grown under a $P_{O_2}$ of 30mTorr, 50mTorr, and 100mTorr by pulsed laser deposition. We found that not only the Co and Ni site-occupations (the Co and Ni compositions) but also their valence states in the films vary depending on the $P_{O2}$. The NCO depositions under lower $P_{O2}$ introduce larger amounts of $Co^{2+}$ and $Ni^{2+}$ into films, influencing their transport and magnetic properties. Furthermore, analyzing XAS and XMCD data quantitatively by the sum rules [20-22], we discuss the origin of the perpendicular magnetic anisotropy in the NCO films.

II. Experimental details

30-nm-thick $Ni_{1-x}Co_{2+y}O_{4-z}$ (NCO) epitaxial films with (001) orientation were fabricated on (100) $MgAl_2O_4$ substrates and under various oxygen pressures ($P_{O2}$s) by pulsed laser deposition. All films studied here is under the substrate-induced compressive strain (0.4%). Previously we showed[19] that when films were grown at a fixed substrate temperature (350 ˚C), the $P_{O2}$ affected the cation distribution (or the cation composition) in the films, influencing transport and magnetic properties. Details of results of basic characterization for grown NCO films, such as X-ray diffraction and magnetization measurements were provided in our previous report[19]. Briefly, by analyzing x-ray diffraction intensities near cations' absorption edges, the $T_d$- and $O_h$-site-occupation of Ni (the amounts of Ni occupying the tetrahedral and octahedral sites) were determined to be 0.17 and 0.86 for the $P_{O2}$=30mTorr film, 0.17 and 0.92 for the $P_{O2}$=50mTorr film, and 0.12 and 0.98 for the $P_{O2}$=100mTorr film. As shown in Fig. 1a, the $T_d$-site-occupation of Ni is almost constant and independent of $P_{O2}$. On the other hand, the $O_h$-site-occupation of Ni increases with increasing $P_{O2}$. In addition, the films grown under the larger $P_{O2}$ have the larger saturated magnetizations and the lower electrical resistivity (Fig. 1b). The ferrimagnetic transition temperature of the films grown under the larger $P_{O2}$ also becomes higher, and the $P_{O2}$ = 100mTorr film becomes a ferrimagnet below 400 K[19].

We carried out XAS and XMCD spectroscopy and characterized the NCO films grown under $P_{O2}$ = 30, 50, and 100 mTorr. Measurements were carried out at the beamlines BL25SU in SPring-8[23,24] and BL-16A in Photon Factory. XAS and XMCD datasets presented in this paper are those obtained from measurements in SPring-8. We note that the dataset acquired from measurements at the two beamlines were essentially the same. The XAS and MCD spectra were recorded at room temperature and under a 1.9 Tesla magnetic field applied at various angles $\theta_H$ with respect to the films' surface. The incident x-ray beam is 10° off the magnetic field direction. The energy resolution in XAS and XMCD spectra is $E/\Delta E=3000$. The XAS spectra for each helicity of the incident beam ($\mu^+$ and $\mu^-$) were obtained in the total electron yield mode and by averaging spectra taken under magnetic fields in opposite directions.

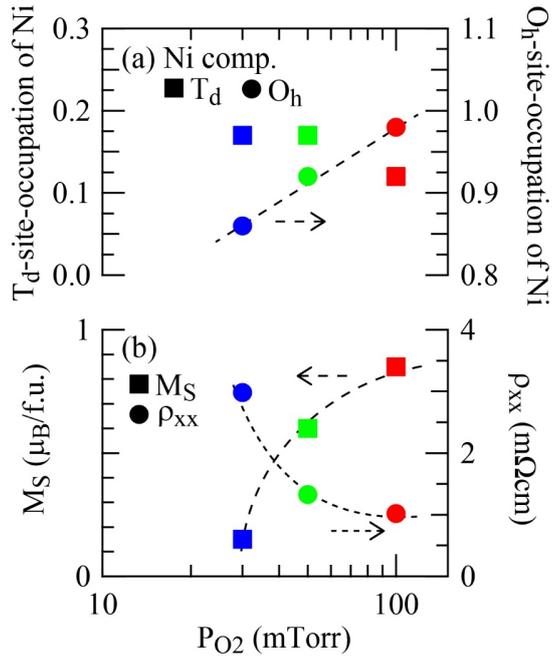

Figure 1: $P_{O2}$ dependence of (a) the tetrahedral ($T_d$) and octahedral ($O_h$) site-occupations of Ni and (b) saturated magnetization and electrical resistivity in NCO epitaxial films. All data in the figures were taken at room temperature. The data in (a) were adopted from Ref. 19. We note that the cation site-occupation (cation composition) in the NCO film is defined as $(Co_{y\text{-Co-Td}}Ni_{x\text{-Ni-Td}})(Co_{y\text{-Co-Oh}}Ni_{x\text{-Ni-Oh}})O_4$. The $T_d$- and $O_h$-site-occupations of cations were determined by assuming that $y_{Co\text{-Td}} + x_{Ni\text{-Td}} = 1$ and and $y_{Co\text{-Oh}} + x_{Ni\text{-Oh}} = 2$.

## III. Results and Discussion

Figures 2a and 2b show the averaged Co and Ni $L_{2,3}$-edge XAS spectra $1/2(\mu^+ + \mu^-)$ for NCO films grown under $P_{O2}$ = 30, 50, and 100mTorr, revealing $P_{O2}$-dependent changes in the valence states of Co and Ni in the films. The data were taken under a 1.9 Tesla magnetic field normal to the films' surface (the $\theta_H = 0°$ configuration). The incident beam is 10° off the magnetic field direction $\theta_H$. In Co $L_3$-edge XAS spectra (Fig. 2a), the pre-edge structure around 777.2eV is pronounced with higher intensity, and the hump structure around 778.8 eV is slightly broader for films grown under lower $P_{O2}$. On the other hand, the peak intensities at 780.2eV are almost constant and independent of $P_{O2}$. Based on reference spectra of Co oxides such as CoO and EuCoO$_3$[18,25,26], the pre-edge structure (around 777.2eV) is characteristic for Co$^{2+}$ octahedrally coordinated by oxygen. The hump (around 778.8eV) arises from both octahedrally and tetrahedrally coordinated Co$^{2+}$. On the other hand, the peak at 780.2eV originates from the Co$^{3+}$ in either tetrahedral or octahedral oxygen coordination. Therefore, the observed $P_{O2}$-dependent changes in Co $L_3$-edge XAS spectra indicates that for the film grown under $P_{O2}$ =100 mTorr and having the cation site-occupation closer to the stoichiometric, the $T_d$- and $O_h$-site Co (referred to as Co_Td and Co_Oh, respectively) dominantly have the +2 and +3 valence states, respectively. We also point out that as revealed by our resonant x-ray diffraction measurements (Fig. 1a), the film grown under the lower $P_{O2}$ has the larger site-occupation of Co_Oh while the Co_Td site-occupation remains almost unchanged against $P_{O2}$. The additionally introduced Co_Oh, which substitutes the $O_h$-site Ni, in the films grown under the lower $P_{O2}$ have the +2 valence states. We note that spectral shape around the Co $L_2$-edge absorption is known to be less dependent on the valence state (+2 or +3). Therefore, the Co $L_2$-edge XAS peak at 794.6 eV exhibit no obvious $P_{O2}$ dependence.

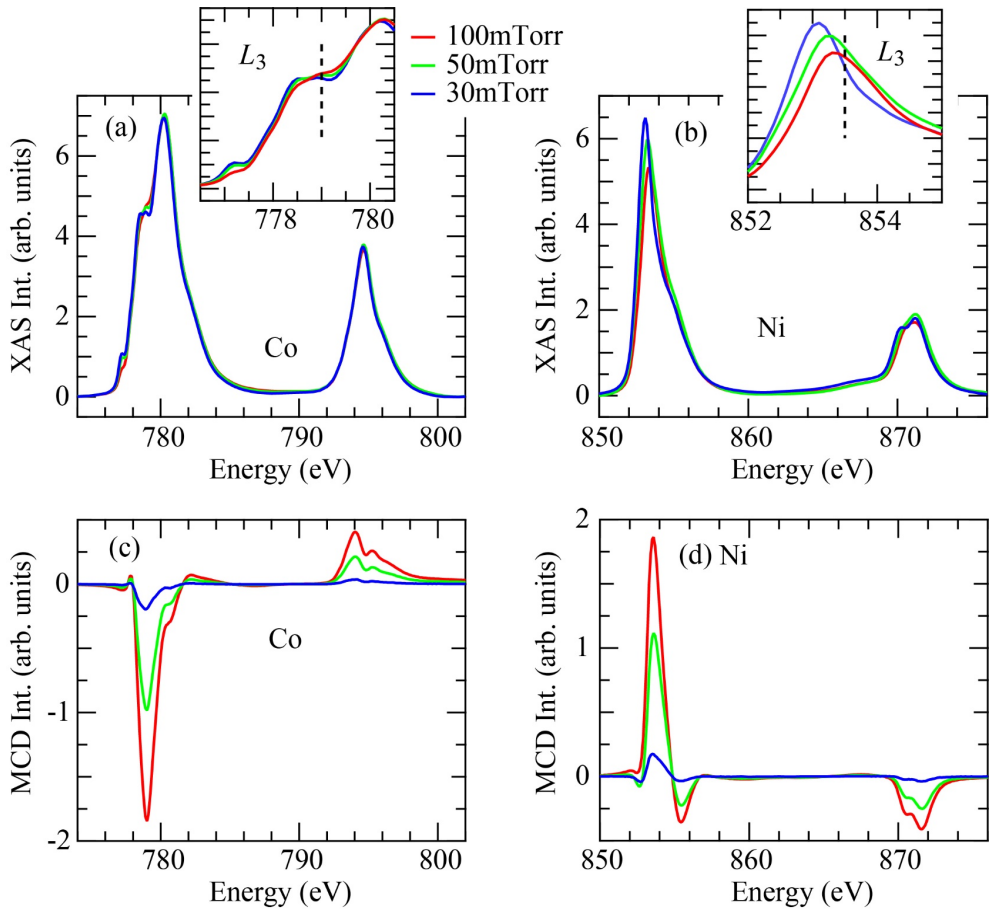

Figure 2: (a, b) Averaged XAS and (c, d) normalized XMCD spectra for NCO films grown under $P_{O2}$ = 30, 50 and 100mTorr. The spectra were in the energy region around (a,c) Co and (b,d) Ni $L_{2,3}$-edge absorptions. The measurements were carried out at room temperature and under a 1.9 Tesla magnetic field normal to the films' surface (the $\theta_H = 0°$ configuration). The incident beam is 10° off the magnetic field direction $\theta_H$. The insets in parts (a) and (b) show expanded views of the Co and Ni-$L_3$ edge absorption peaks.

The Ni $L_{2,3}$-edge absorption spectra in Figure 2b also depend on $P_{O2}$. The peak positions of the Ni $L_3$-edge absorptions for the films grown under higher $P_{O2}$ shift toward the higher-energy side. Concomitantly, the shoulder structures associated with the Ni $L_2$-edge absorptions (around 870.3eV) are less pronounced for the larger-$P_{O2}$ films. These observations indicate that Ni in the $O_h$-site is in the intermediate valence state between +2 and +3, Ni$^{(2+\delta)+}$ ($0<\delta<1$). We note that the $L_3$-edge peak position for the $P_{O2}$ = 30mTorr film (853.1 eV) is almost identical to the $L_3$-edge peak position characteristic of the Ni$^{2+}$ octahedrally coordinated by oxygen (~853 eV) [18,27,28]. In addition, the $O_h$-site-occupation of Ni in the film decreases with decreasing $P_{O2}$ [19]. The $P_{O2}$-dependent shifts of the Ni $L_3$-edge absorption peaks, therefore, indicate that growing films under lower $P_{O2}$ leads not only to the decrease in the $O_h$-site-occupation of Ni but also to the lowering of its valence state. The Ni valence state for the $P_{O2}$ = 30mTorr film is closer to +2. By reproducing the experimentally observed Ni XAS spectra based on reference spectra of $Ni^{2+}$ and $Ni^{3+}$ $L$-edge absorptions, the Ni nominal valence state for the $P_{O2}$ = 100mTorr film is estimated to be about +2.5 ($\delta \sim 0.5$) [29]. The observed changes in the valence states of Co and Ni imply that the oxygen contents in NCO films also vary depending on $P_{O2}$ and that larger amounts of oxygen vacancies are accommodated in films grown under lower $P_{O2}$[29].

The $P_{O2}$-dependent changes in the Co and Ni valence states impact on magnetic properties as revealed by XMCD spectroscopy. Figures 2c and 2d respectively show the Co and Ni $L_{2,3}$-edge XMCD spectra for the $P_{O2}$ = 30, 50, and 100mTorr films. The XMCD signals ($\Delta\mu = \mu^+ - \mu^-$) are observed in both Co and Ni absorption edges for all the films. While the signal intensities depend on $P_{O2}$, the whole spectral shapes and the peak positions remain almost unchanged independent of $P_{O2}$. Both Co and Ni XMCD peaks for the films grown under larger $P_{O2}$ have higher intensities, indicating that both spin magnetic moments of Co and Ni are larger for films having the larger magnetization.

Interestingly, the Co $L_3$-edge XMCD signals are enhanced at 779 eV where the x-ray absorptions of the $T_d$- and $O_h$-site $Co^{2+}$ occur, and their magnitudes become larger for the films grown under the larger $P_{O2}$ whose $O_h$-site-occupation of $Co^{2+}$ is smaller. In addition, the $O_h$-site $Co^{3+}$ should be in the low spin state with S = 0 [1,4,18], and no contribution of Co_Oh to the XMCD signal is expected. Therefore, the observed $P_{O2}$ dependence of the Co XMCD signal implies that the spin magnetic moments in the $T_d$-site $Co^{2+}$ (S = 3/2) dominantly contribute to the signal. The Ni $L_3$-edge XMCD signal in Fig. 2d also becomes larger for the films grown under the larger $P_{O2}$ and having the $O_h$-site Ni with the higher intermediate valence state. While the observed spectral shape of the Ni XMCD signals resembles that characteristic of octahedrally coordinated $Ni^{2+}$ [30], it is unlikely that the $Ni^{2+}$ in the $O_h$-site, which has the S = 1 configuration, give rise to XMCD signal. The $O_h$-site-occupations of $Ni^{2+}$ is larger for the films grown under the lower $P_{O2}$ while their Ni XMCD signals are largely suppressed (Fig. 2d). We also note that the energy positions of Ni XMCD signals are constant independent of $P_{O2}$, which is in contrast to the $P_{O2}$-dependent variations of the nominal valence of Ni (Figure 2b). These observations indicate that the Ni having the intermediate valence state, $Ni^{(2+\delta)+}$ is the key for the spin magnetic moment. The electronic structure of $Ni^{(2+\delta)+}$ that gives rise to the XMCD signal differs from those of $Ni^{2+}$ and $N^{3+}$.

Our XAS and XMCD results highlight the significance of the $T_d$-site Co (Co_Td) and $O_h$-site Ni (Ni_Oh) as the keys that determine the magnetic and transport properties of NCO films. Recent first-principles calculations showed that the density of states at the Fermi level in stoichiometric NCO consists of spin-down electrons of the Ni_Oh, leading to the half-metallic electronic structure [6,7]. Importantly, when the $O_h$-site Ni has the intermediate valence state (higher than +2), its $e_g$ orbitals are partially unoccupied, and conduction carriers are provided. Given the half-metallic electronic structure in NCO, the delocalization of the conduction carriers should preferably align the spin magnetic moments in each $O_h$ and $T_d$ sub-lattice. It is worth pointing out that reducing the Ni site-occupation and lowering its valence state results in localization of the conduction carriers, and thus the spin magnetic moments in the films with lowered electrical conduction are more difficult to be aligned. Therefore, the films grown under the larger $P_{O2}$ have the lower electrical resistivity and the larger magnetization (Fig. 1).

We also note that the signs of the Co and Ni XMCD signals are opposite each other, which is further confirmed from the magnetic field dependences of Co and Ni $L_3$-edge XMCD signals in Figure 3. For all films, the Co and Ni signals in the positive magnetic field regions are negative and positive, respectively, indicating that the spin magnetic moments of the Co_Td and Ni_Oh align in the anti-parallel manner. The hysteresis behavior of the signals against the magnetic field sweep direction becomes prominent for the films grown under the larger $P_{O2}$, and the remnant values of both Co and

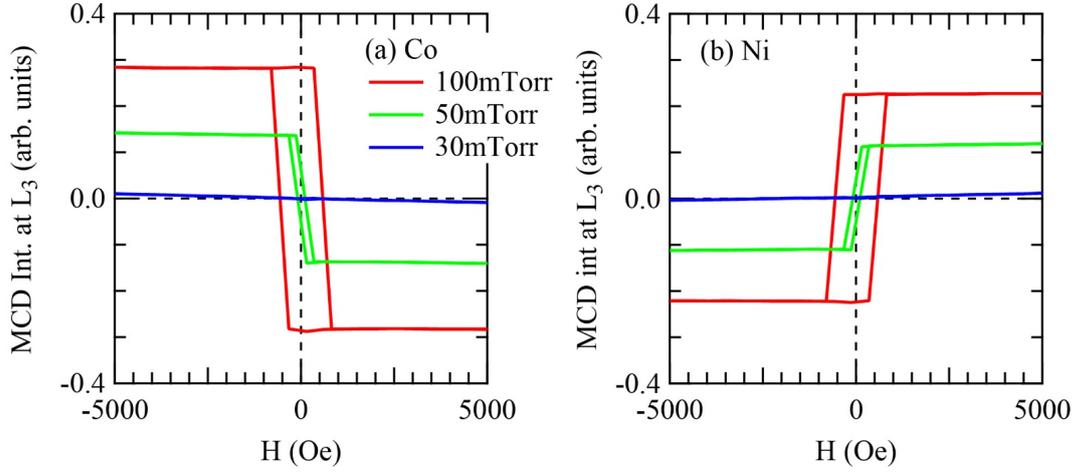

Figure 3: Magnetic field dependence of (a) Co $L_3$- and (b) Ni $L_3$-XMCD peak intensity for NCO films grown under $P_{O2}$ = 30, 50 and 100mTorr. The data were collected in the $\theta_H$=0° configuration and with the incident x-ray energy fixed to be 779.0eV for the Co $L_3$-XMCD signal and 853.5eV for the Ni signal.

Ni signals at zero magnetic field also become larger. These observations are in close agreement with the fact that the films grown under the larger $P_{O2}$ have the larger magnetization and the higher ferrimagnetic transition temperature [19].

We further evaluate spin and orbital magnetic moments originating from the Ni_Oh and Co_Td quantitatively and investigate how these cations contribute to the orbital magnetic moments that determine the perpendicular magnetic anisotropy in the films. Figure 4 shows the Co and Ni $L_{2,3}$-edge XMCD spectra and their integrated ones obtained with the $\theta_H$ = 0°, 40°, and 80° configurations. The inset of the figure schematically shows the measurement configuration including the magnetic field angle $\theta_H$. The Co $L_3$-edge XMCD peak intensities and the integrated XMCD signals decrease with increasing $\theta_H$. In contrast, while the Ni XMCD signals and their integrated intensities slightly vary against the change in $\theta_H$, no clear relationships between the signal variations and $\theta_H$ are seen. Given that the value obtained by integrating the $L_2$ and $L_3$ XMCD peaks is proportional to the orbital magnetic moment according to the sum rules [20-22], the $\theta_H$ dependence observed in the Co and Ni XMCD signals implies that the orbital magnetic moment of Co is anisotropic whereas that of Ni is rather symmetric, indicating that the Co plays the dominant role in determining the magnetic anisotropy in NCO.

To quantitatively evaluate orbital and spin magnetic moments, we applied the sum rules to the XAS and XMCD spectra obtained at various $\theta_H$ and calculated the moments. We note that the Co XMCD signals mainly originates from the Co_Td dominantly having the +2 valence state while none of the Co_Oh contribute to the magnetic moment. The number of electrons in 3d orbitals, $n_{3d}$, for Co is thus assumed to be 7 (corresponding to $Co^{2+}$). Taking into account that the energy position of the Ni $L_3$-edge XMCD peak is close to that of the Ni $L_3$-edge absorption peak for the $P_{O2}$=100mTorr film as shown in the inset of Figure 2b, we assume that the Ni valence state that dominantly contributes to the XMCD signal is +2.5 and the $n_{3d}$ for Ni is 7.5. Figure 5 shows the $\theta_H$ dependence of the spin and orbital magnetic moments of Co_Td and Ni_Oh in the NCO films grown under $P_{O2}$ = 30, 50, and

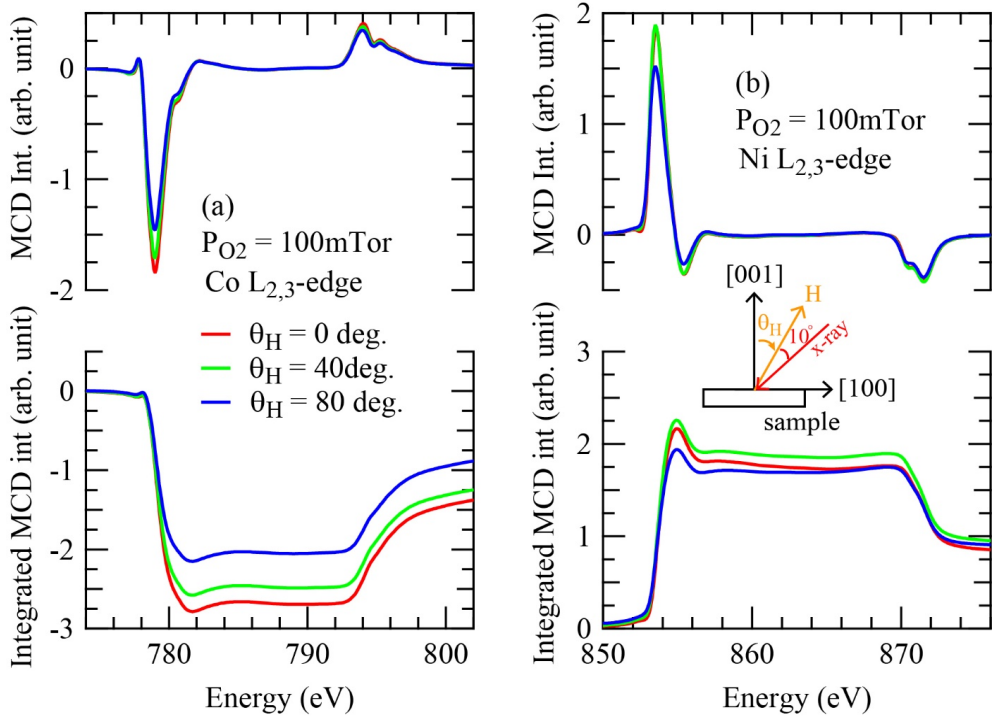

Figure 4: (a) Co $L_{2,3}$- and (b) Ni $L_{2,3}$-XMCD spectra and their integrated intensity for the $P_{O2}$ = 100mTorr film. The spectra were measured with the magnetic field angle $\theta_H$ of 0, 40, and 80 degrees. The measurement configuration is shown in the inset of the figure.

100mTorr. As expected from the $P_{O2}$-dependence of the XMCD signals (Figs. 2 and 3), the spin magnetic moments of both Co_Td and Ni_Oh in the films grown under the larger $P_{O2}$ become larger. Both Co_Td and Ni_Oh spin magnetic moments are $\theta_H$-dependent with their maximums at $\theta_H = 0°$. In addition, the orbital magnetic moments of Co_Td and Ni_Oh are $P_{O2}$-dependent, and their trends follow those of the $P_{O2}$-dependence of the spin magnetic moments. For both Co_Td and Ni_Oh, the orbital and spin moments have the same sign. Thus, the orbital and spin moments are parallel aligned in each $T_d$ and $O_h$ sub-lattice. Importantly, the Co_Td orbital magnetic moment is $\theta_H$-dependent and is maximized at $\theta_H = 0°$ while the Ni_Oh orbital magnetic moment is almost $\theta_H$-independent. For the $P_{O2} = 100$mTorr film whose perpendicular magnetic anisotropy energy is largest among the films investigated in this study [19], the Co_Td orbital moment is as large as $0.14\mu_B$/Co_Td while the Ni_Oh one is as small as 0.07 $\mu_B$, indicating that the Co_Td play the dominant role in determining the perpendicular anisotropy and that the contribution of the Ni_Oh to the magnetic anisotropy is less dominant. This observation is in agreement with the recent theoretical investigation [10] showing that the out-of-plane orbital magnetic moment originates from the $d_{x^2-y^2}$ orbital in the $T_d$-site Co, leading to the perpendicular magnetic anisotropy in NCO. Our results indicate the significance of the site-dependent cation valence states for the magnetic and transport properties of $NiCo_2O_4$ films. The Co_Td determines the magnetic anisotropy while it provides almost no density of states at the Fermi level and less influence the magnetization and electrical conduction of the NCO films. On the other hand, the Ni_Oh has little contribution to the magnetic anisotropy, while it closely ties with the electronic structure around $E_F$, strongly influencing the magnetization and the electorical conduction. These correlations between the site-dependent cation states and the functional proerpties explain why $NiCo_2O_4$ has the half-metallic band structure and the perpendicular magnetic anisotropy simultaneously.

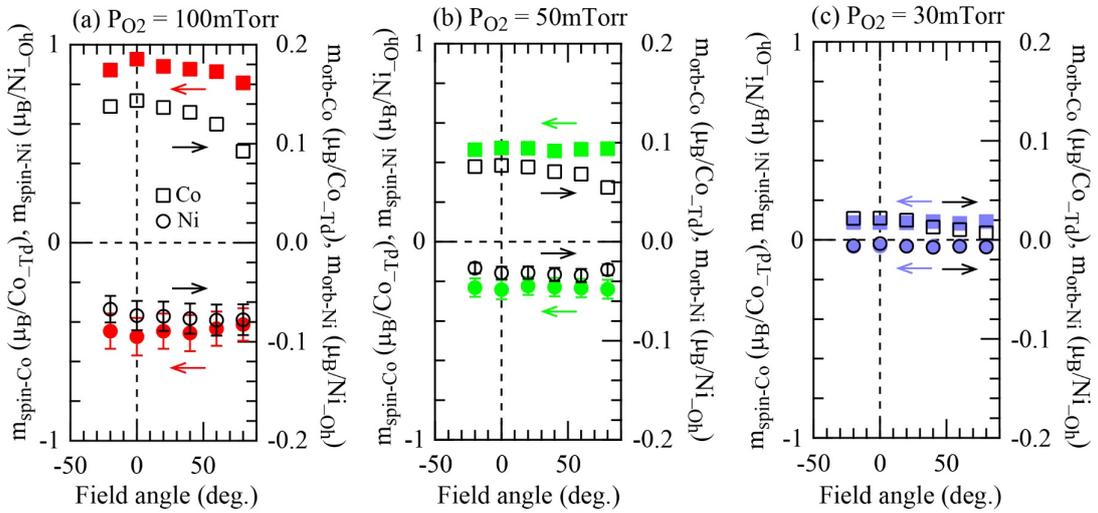

Figure 5: Magnetic field angle dependence of spin and orbital magnetic moments ($m_{spin}$ and $m_{orb}$) originating from the $T_d$-site Co (Co_Td) and the $O_h$-site Ni (Ni_Oh) in NCO films grown under $P_{O2}$ = (a) 100, (b) 50 and (c) 30mTorr. The filled squares and circles are for the $m_{spin}$ from the Co_Td and Ni_Oh, respectively. The open squares and circles are for the $m_{orb}$ from the Co_Td and Ni_Oh, respectively. Errors in the spin and orbital magnetic moments of Ni_Oh are estimated from the possible variation of $n_{3d}$, i.e. $7 < n_{3d} < 8$. In (c), the errors in the Ni moments are not included in the figure because they are relatively small (smaller than the marker sizes)

IV. Summary

We evaluated the cation valence states and spin and orbital magnetic moments in the ferrimagnetic $Ni_{1-x}Co_{2+y}O_{4-z}$ (NCO) epitaxial films with the perpendicular magnetic anisotropy. We found that the oxygen pressure $P_{O2}$ during the film growth by pulsed laser deposition influences not only the cation stoichiometry but also the cation valence states. For the film having a cation site-occupation close to the stoichiometric one, the $O_h$-site Ni is in the intermediate valence state while the $T_d$-site and $O_h$-site Co dominantly have the +2 and +3 valence states, respectively. On the other hand, for the films grown under the lower $P_{O2}$ the $O_h$-site-occupation of Ni is reduced, and its valence state is also lowered closer to +2. The Co additionally accommodated in the $O_h$-site is in the +2 valence state while the valence state of the $T_d$-site Co remains unchanged. The spin magnetic moments in the NCO films dominantly originate from the $T_d$-site Co (Co_Td) and $O_h$-site Ni (Ni_Oh). Both Co_Td and Ni_Oh spin moments become larger when the cation site-occupation is closer to the stoichiometric one. These observations indicate that the intermediate valence state of the Ni_Oh is responsible for the delocalization of conduction carriers and plays the key role in stabilizing the electrical conduction and the ferrimagnetic moment. We also show that the orbital magnetic moment originating from the Co_Td is as large as 0.14 $\mu_B$/Co_Td and is anisotropic, parallel to the out-of-plane direction. On the other hand, the Ni_Oh orbital moment is as small as ~0.07 $\mu_B$/Ni_Oh and is rather isotropic. The Co_Td therefore plays the key role in the perpendicular magnetic anisotropy in the films. Our results demonstrate the significance of the site-dependent cation valence states for the magnetic and transport properties of $NiCo_2O_4$ films.

Acknowledgment

This work was partially supported by a grant for the Integrated Research Consortium on Chemical Sciences, by Grants-in-Aid for Scientific Research (Grant Nos. JP16H02266, JP17H04813, JP18K141130 and JP19H05816), by a JSPS Core-to-Core program (A), and by a grant for the Joint Project of Chemical Synthesis Core Research Institutions from the Ministry of Education, Culture, Sports, Science and Technology (MEXT) of Japan. The synchrotron radiation experiments at Photon Factory were performed with the approval of the Photon Factory Program Advisory Committee (proposal No. 2019PF-29). The XAS and XMCD measurements at SPring-8 were made with the approval of the Japan Synchrotron Radiation Research Institute (Proposal No. 2019B1266).